\documentclass[aps,prl,twocolumn,superscriptaddress,longbibliography]{revtex4-2}
\usepackage{breakurl}
\usepackage{xcolor}
\usepackage{sidecap}
\usepackage{amssymb}
\usepackage{hhline}
\usepackage{multirow}

\sidecaptionvpos{figure}{t}
\usepackage{amsmath}
\usepackage{mathrsfs}
\usepackage{graphicx}
\usepackage{esint}
\usepackage{epstopdf}
\usepackage{rotating}
\graphicspath{{pict/}{./}}
\usepackage{bm}%
\usepackage{microtype,bm,bbm,graphicx,booktabs,times}

\newcounter{Fig}

\newcommand\mymapstol{\mathrel{\ooalign{$\leftarrow$\cr%
  \kern1.75ex\raise0.275ex\hbox{\scalebox{1}[0.4]{$\mid$}}\cr}}}

\newcommand\mymapstor{\mathrel{\ooalign{$\rightarrow$\cr%
  \kern-.15ex\raise.275ex\hbox{\scalebox{1}[0.4]{$\mid$}}\cr}}}
  
\usepackage{graphicx,tipa}

\begin{document}

\title{Polarizations Underdescribe Vectorial Electromagnetic Waves}
\author{Chunchao Wen}
\author{Jianfa Zhang}
\email{jfzhang85@nudt.edu.cn}
\author{Chaofan Zhang}
\email{c.zhang@nudt.edu.cn}
\author{Shiqiao Qin}
\author{Zhihong Zhu}
\author{Wei Liu}
\email{wei.liu.pku@gmail.com}
\affiliation{College for Advanced Interdisciplinary Studies, National University of Defense Technology, Changsha 410073, P. R. China.}
\affiliation{Nanhu Laser Laboratory and Hunan Provincial Key Laboratory of Novel Nano-Optoelectronic Information Materials and
Devices, National University of Defense Technology, Changsha 410073, P. R. China.}

\begin{abstract}

Electromagnetic waves are described by not only polarization ellipses but also cyclically rotating vectors tracing out them. The corresponding fields are respectively directionless steady line  fields and directional instantaneous vector fields. Here we study the seminal topic of electromagnetic scattering from the perspective of instantaneous vector fields and uncover how the global topology of the momentum sphere regulates local distributions of tangent scattered fields. Structurally-stable generic singularities of  vector fields  move cyclically along lines of linear polarizations and at any instant the index sum of all instantaneous singularities has to be the Euler characteristic $\chi=2$. This contrasts sharply with steady line fields, of which generic singularities constrained by the Euler characteristic locate on points of circular polarizations. From such unique perspective of instantaneous singularities, we discovered that for circularly-polarized waves scattered by  electromagnetic duality-symmetric particles, since linearly-polarized scatterings are prohibited by helicity conservation, there must exist at least one dark direction along which the scattering is strictly zero. Two such dark directions can be tuned to overlap, along which the scattering would remain zero for arbitrary incident polarizations. We have essentially revealed that \textit{polarizations underdescribe vectorial electromagnetic waves and the instantaneous perspective is indispensable}. The complementarity we discover provides broader and deeper insights into not only electromagnetism, but also other branches of wave physics where singularities are generic and ubiquitous.
\end{abstract}

\maketitle

\begin{figure}[tp]
\centerline{\includegraphics[width=8cm]{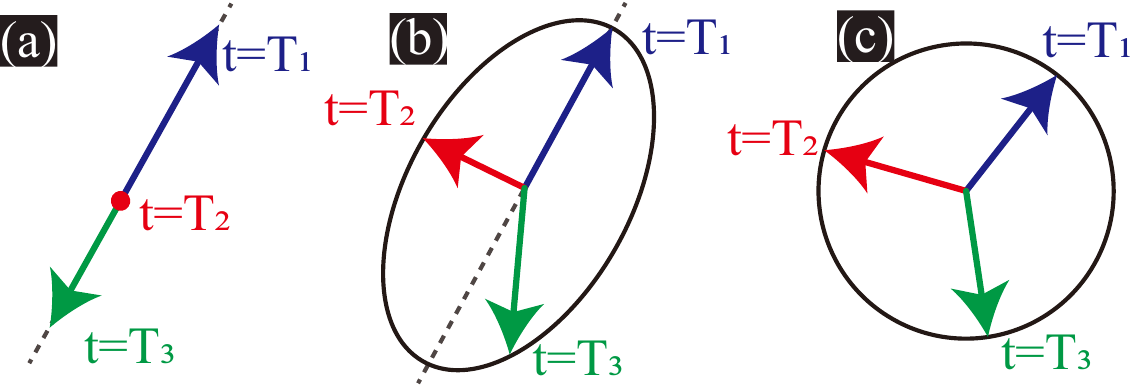}} \caption{\small Linear, elliptic and circular polarizations are shown in (a)-(c), respectively.  The line fields (dashed lines) is defined as the orientation line of linear poalrization in (a), the major axis of the ellipse in (b), and ill-defined in (c). Instantaneous field vectors  of which the tips trace out the polarization ellipses are shown at three instants ($\mathrm{t=T_{1,2,3}}$) for each polarization. \textit{Only the linear polarization accomodate zero instantaneous field vectors (red dot) at the instant $\mathrm{t=T_{2}}$}.}
\label{fig0}
\end{figure}

Maxwell equations are expressed by time-varying vector fields and for monochromatic waves instantaneous field vectors oscillate cyclically both in space and time~\cite{JACKSON_1998__Classical}. The oscillation speed of optical (or higher-frequency) waves is orders of magnitude higher than that of currently available detectors, and thus observable physical quantities up-to-date have been essentially characterized by time-averaged parameters (\textit{e.g.} the term ``\textit{time-averaged}'' frequently appears in most chapters of the classic book~\cite{JACKSON_1998__Classical}). To some extent, polarization ellipses (trajectories of cyclically-rotating instantaneous vectors' tips) are also time-averaged steady structures that are widely employed for descriptions of electromagnetic polarizations~\cite{JACKSON_1998__Classical,YARIV_2006__Photonics} (refer to Fig.~\ref{fig0}). Their singularities of linear and circular polarizations (normal and orientation directions of ellipses are undefined, respectively) are central concepts in singular optics~\cite{NYE_natural_1999}, currently bridging different vibrant disciplines of photonics~\cite{GBUR_2016__Singular,CHEN_Proc.Natl.Acad.Sci._Evolution,LIU_ArXiv201204919Phys._Topological,KOSHELEV_2019_ScienceBulletin_Metaoptics,KANG_2023_NatRevPhys_Applications,WANG_2024_PhotonicsInsights_Optical}.

Considering that direct observations of high-frequency field oscillations are currently inaccessible and thus time-averaged steady quantities and structures are sufficient for descriptions of observable effects, does it mean that the perspective of instantaneous field vectors is superfluous and will not be able to provide new physical insights? Here we reveal that \textit{polarizations underdescribe the vectorial nature of electromagnetic waves and an extra instantaneous perspective is indispensable}. We study the seminal topic of electromagnetic scattering from the perspective 
 of  Poincaré-Hopf theorem (of which a reduced scenario is the hairy ball theorem)~\cite{RICHESON_2012__Euler,NEEDHAM__Visuala}, and unveil how the global topology of the scattering momentum sphere (topological Euler characteristic $\chi=2$) regulates the local distributions of the scattered fields. In the far field, the scattered waves are transverse, which are described by tangent fields on the momentum sphere: both steady directionless line fields (vectors without arrows; defined as the major or minor axes of the polarization ellipses) and cyclically oscillating directional vector fields (refer to Fig.~\ref{fig0}). We have previously revealed that throughout the momentum sphere the index sum of line-field singularities of circular polarizations has to be the Euler characteristic $\chi=2$~\cite{CHEN_2019__Singularities,CHEN_2019_ArXiv190409910Math-PhPhysicsphysics_Linea,CHEN_2020_ACSOmega_Global,Supplemental_Material}. Here we apply the same theorem to instantaneous vector fields, and reveal that:  structurally-stable generic singularities (instantaneous vectorial zeros) always move along lines of linear polarizations in a cyclic manner; they are irrelevant to and never locate at points of circular or elliptic polarizations (see Fig.~\ref{fig0});  their index sum also has to be $2$, at any instant.  

From this perspective of instantaneous singularities and parameter-space topology, we examine scatterings by electromagnetic duality-symmetric particles~\cite{FERNANDEZ-CORBATON_2013_Phys.Rev.Lett._Electromagnetica,FERNANDEZ-CORBATON_2013_Opt.ExpressOE_Forwarda,YANG_2020_ACSPhotonics_Electromagnetic}. For circularly-polarized (CP) incident waves, helicity conservation prohibits linearly-polarized scatterings and thus instantaneous singularities can only locate on positions where the field is zero at any instant: there must be at least one dark direction (\textbf{V} point) of index $2$ where the scattering (time-averaged intensity $I_{\mathrm{sca}}$) is zero. We further show when a pair of \textbf{V} points for incident left-handed and right-handed CP (LCP and RCP) waves are tuned to overlap, this direction will remain dark for arbitrary incident polarizations. Though instantaneous singularities are generally not experimentally observable,  without such unique insight of instantaneity the protected existence of duality-induced dark directions cannot possibly be revealed. Deeper understanding and thorough exploitations of electromagnetic effects require both instantaneous and time-averaged steady quantities and structures. Our discovery can be extended to other wave physics disciplines where singularities would generically emerge.

For electromagnetic scattering by finite scattering bodies of arbitrary geometric and optical parameters, scattered far fields are transverse and automatically tangent on the momentum sphere. For the associated line fields  on the sphere, the Poincaré-Hopf theorem requires that~\cite{NYE_natural_1999,NEEDHAM__Visuala,Supplemental_Material}:
\begin{equation}
\label{PHtheorem-line}
\sum_i \operatorname{{\mathbf{Ind}}}({\mathrm{\mathbf{C}}}_i)=\chi=2,  
\end{equation}
where $\mathrm{\mathbf{C}}_i$ are isolated line-field singularities and their generic (structurally stable against perturbations) forms correspond to CP scatterings with indexes (Hopf index) ${\mathrm{{\mathbf{Ind}}}}=\pm 1/2$. Non-generic $\mathrm{\mathbf{C}}_i$ (higher-order CP scatterings or \textbf{V} points of other half-integer indexes)  would be broken into pairs of generic CP scatterings upon perturbations. It is worth mentioning that though linear polarizations are also polarization singularities, they are not directly related to the 
Poincaré-Hopf theorem in Eq.~(\ref{PHtheorem-line}). \textit{If throughout the momentum sphere there are no CP scatterings, then according to Eq.~(\ref{PHtheorem-line}) there must be at least one direction of zero scattering (non-generic polarization singularity; \textbf{V} point) and the index sum of the zero(s) must be $2$}. A special such scenario (scatterings have linearly-polarized) has been demonstrated in our previous study~\cite{CHEN_2019__Singularities}.

We proceed to apply the Poincaré-Hopf theorem  to the instantaneous vector (electric or magnetic) fields on the momentum sphere~\cite{NYE_natural_1999,NEEDHAM__Visuala,Supplemental_Material}:
\begin{equation}
\label{PHtheorem-vector}
\sum_i \operatorname{{\mathbf{Ind}}}({\mathrm{\mathbf{Z}}}_i)=2.  
\end{equation}
Here $\mathrm{\mathbf{Z}}_i$ are isolated vector-field singularities located at the position where the instantaneous field is zero and thus the vector direction is not defined. $\mathrm{\mathbf{Z}}_i$ and $\mathrm{\mathbf{C}}_i$  differ in the following aspects: \textbf{(i)}  $\mathrm{\mathbf{Z}}_i$ are singularities of instantaneous vector fields and thus they can move in time (except for non-generic forms of \textbf{V} points), while  $\mathrm{\mathbf{C}}_i$ are singularities of steady line fields and thus they are spatially fixed; \textbf{(ii)} Vector fields are directional (return to itself with at least 2$\pi$ rotation) and thus  $\mathrm{\mathbf{Z}}_i$ have integer indexes (generic singularity indexes are $\pm 1$), while line fields are directionless ($\pi$ rotation is sufficient for the return) and thus $\mathrm{\mathbf{C}}_i$ have half-integer indexes~\cite{Supplemental_Material}; \textbf{(iii)} Generic $\mathrm{\mathbf{Z}}_i$ move along lines of linear polarizations in a cyclic manner while generic $\mathrm{\mathbf{C}}_i$ locates at CP positions. The last aspect is apparent (see Fig.~\ref{fig0}): tips of instantaneous field vectors of  CP (or elliptically-polarized) waves trace out circles (or ellipses) and thus they can never reach zero at any instant; only linear polarizations (field vectors can read zero at discrete instants) accommodate vector-field singularities. 

Generally speaking, $\mathrm{\mathbf{C}}_i$ and $\mathrm{\mathbf{Z}}_i$ are different points locating on different positions of circular and linear polarizations, respectively. The only exception is that they overlap at \textbf{V} points which are singularities of singularities: a \textbf{V} point can be viewed both as a special $\mathrm{\mathbf{C}}_i$ point in Eq.~(\ref{PHtheorem-line}) and as a special $\mathrm{\mathbf{Z}}_i$ point in Eq.~(\ref{PHtheorem-vector}).  Based on the argument of instantaneous vector fields, we can reach a similar conclusion as that for line field: \textit{if throughout the momentum sphere there are no linearly-polarized scatterings, then according to Eq.~(\ref{PHtheorem-vector}) there must be at least one direction of zero scattering and index sum of the zero(s) must be $2$}. From the joint perspective of both line and vector fields we can reach the following conclusion: \textit{{scattering (radiation) patterns without dark directions must have both points of circular polarizations and lines of linear polarizations; or equivalently, patterns without points of circular polarizations or lines of linear polarizations must have dark directions}}. Our arguments above are fully based on the Poincaré-Hopf theorem and thus the conclusions drawn are universally applicable to all scatterings scenarios for scatterers of arbitrary geometric and optical parameters and for arbitrarily-structured incident waves. 

\begin{figure}[tp]
\centerline{\includegraphics[width=8cm]{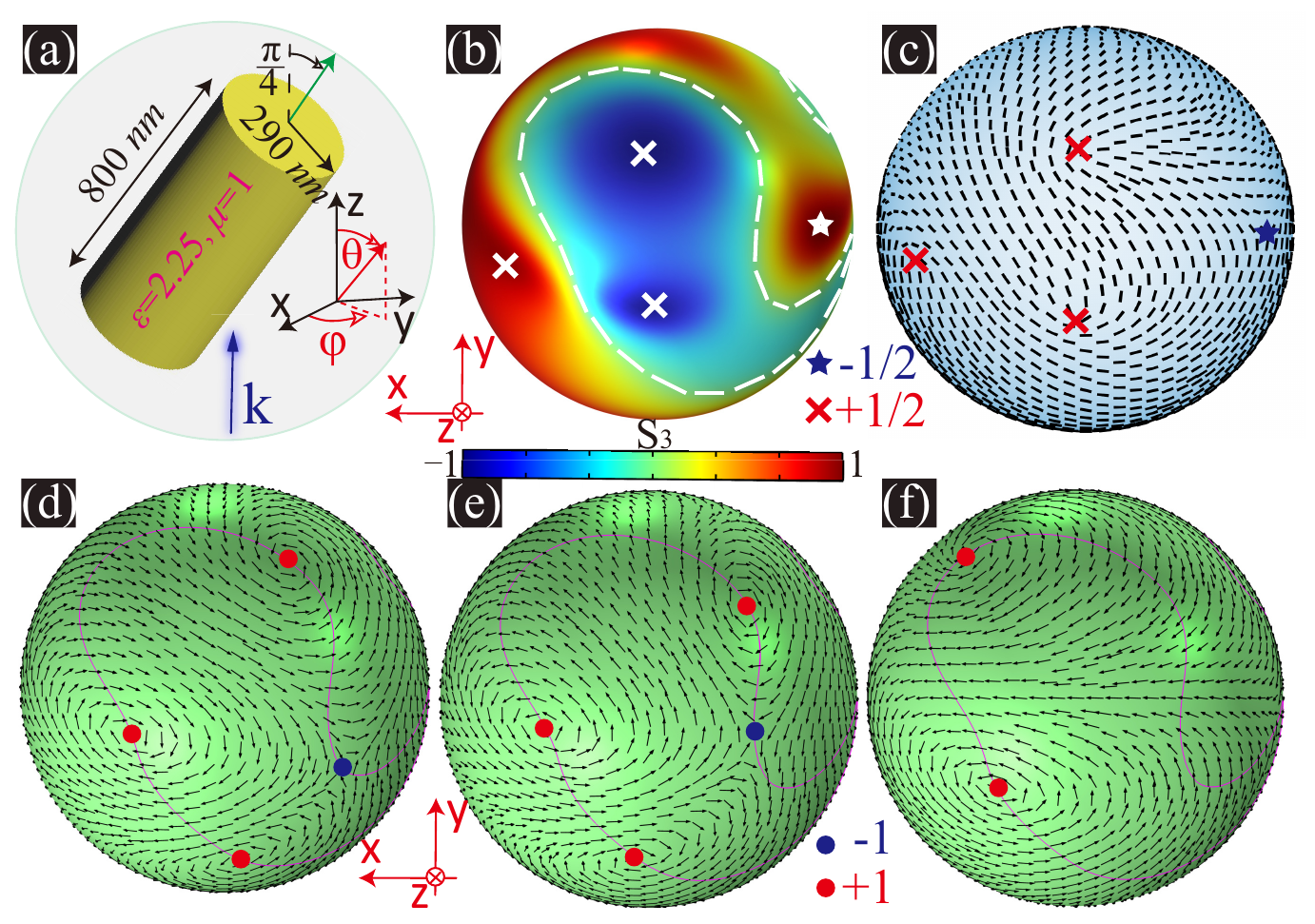}} \caption{\small (a) The scattering of a dielectric cylinder by plane waves incident along +\textbf{z}. (b) Polarization ($S_3$) distributions of scattered fields on the momentum sphere, with four in-view \textbf{C} points marked and indexes specified. Lines of linear polarizations are also marked by curves. The corresponding line fields (c) and instantaneous vector fields at three different instants (d-f) are also shown, with in-view \textbf{Z} points marked and their indexes specified.}
\label{fig1}
\end{figure}

 Now we exemplify our conclusions  through an elementary scattering configuration (numerical results are obtained through COMSOL MULTIPHYSICS) shown in Fig.~\ref{fig1}(a): a dielectric cylinder (relative permittivity $\epsilon=2.25$ and permeability $\mu=1$) scatters LCP plane wave of incident wavelength $\lambda$ and wavevector \textbf{k} ($k$=$|$\textbf{k}$|$). The far-field polarization distributions (in terms of the Stokes parameter $S_3$~\cite{YARIV_2006__Photonics}; $S_3=\pm 1$ corresponds respectively to LCP and RCP states; $\lambda=810$ nm) are shown in Fig.~\ref{fig1}(b), where four in view line-field singularities of circular polarizations (\textbf{C} points) are marked with their indexes specified.  In  Fig.~\ref{fig1}(b) we have also marked (by dashed curves) lines of linear polarizations ($S_3=0$). The corresponding line field (semi-major axes of polarization ellipses) distributions are shown in Fig.~\ref{fig1}(c), manifesting clearly both positions and indexes of the \textbf{C} points. In addition, there are $2$ other \textbf{C} points out of view, with the index sum of all \textbf{C} points being exactly $2$~\cite{Supplemental_Material}.

Meanwhile we also show the instantaneous electric field vector distributions on the momentum sphere at three instants in Figs.~\ref{fig1}(d)-\ref{fig1}(f). In contrast to steady line fields, vector field singularities (instantaneous vector zeros; \textbf{Z} points) cyclically move  along lines of linear polarizations (also marked), with field patterns repeating themselves in every cycle $\mathrm{T}=\lambda/c$ ($c$ is the speed of light)~\cite{Supplemental_Material} (more details about the birth and annihilation of opposite-index singularities on lines of linear polarizations could be found in the supplemental video). In Figs.~\ref{fig1}(d)-\ref{fig1}(f) only in-view \textbf{Z} points are marked with their indexes specified  and the index sum of all \textbf{Z} points are indeed $2$~\cite{Supplemental_Material}.

\begin{figure}[tp]
\centerline{\includegraphics[width=8cm]{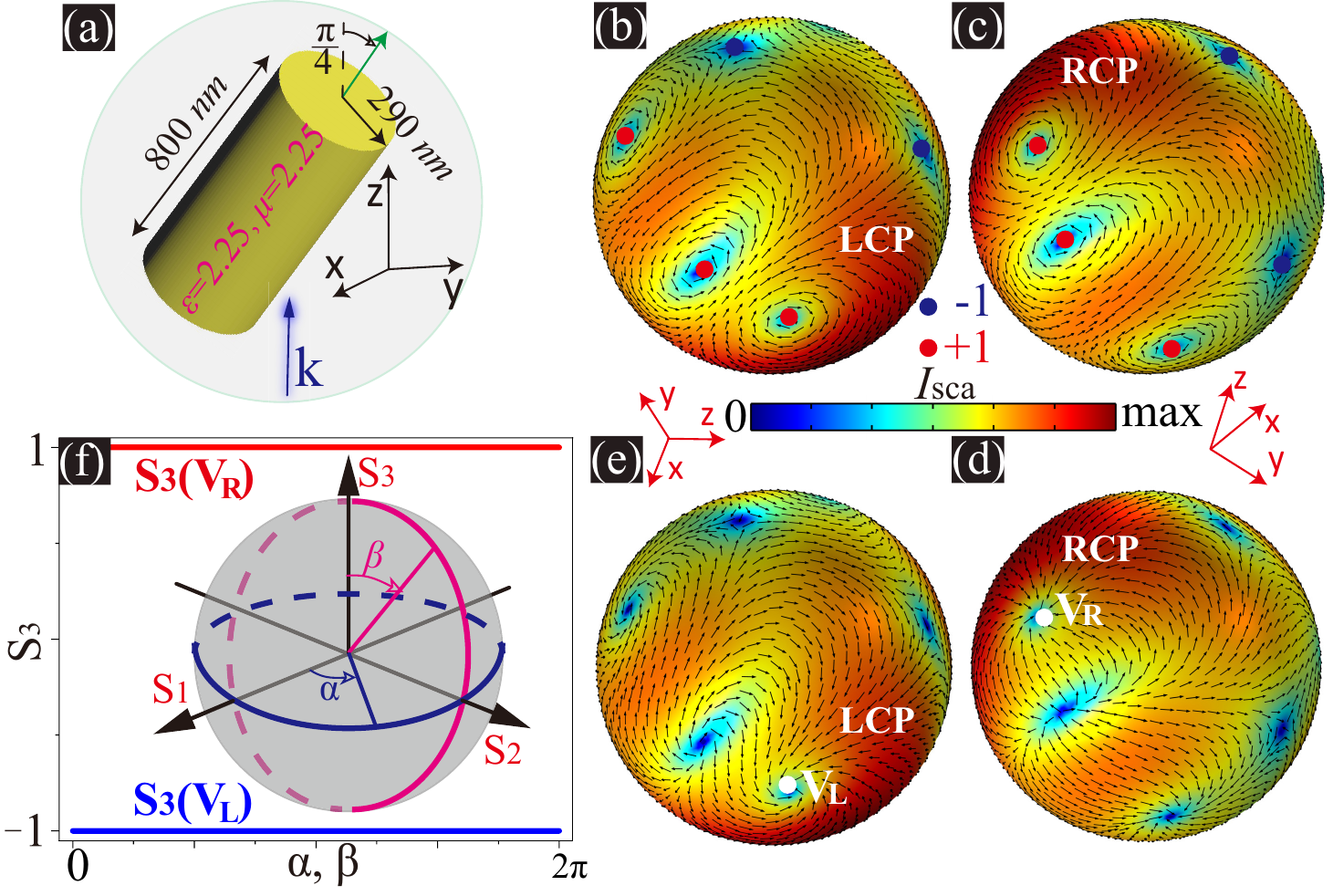}} \caption{\small (a) The scattering of plane waves along +\textbf{z} by a self-dual cylinder. (b) \& (c)  $I_{\mathrm{sca}}$ (color bar) and instantaneous field vector distributions respectively for LCP and RCP incidences, with in-view \textbf{V} points marked and their indexes specified. (d) \& (e) are the same as  (b) \& (c), but at a different instant, where two \textbf{V} points ($\mathbf{V}_{\mathrm{L}}$ and $\mathbf{V}_{\mathrm{R}}$) are marked. (f) The invariance  of scattering polarizations on these marked points, with the inset showing the Poincaré sphere on which two great circles are parameterized by $\alpha$ and $\beta$, respectively.}
\label{fig2}
\end{figure}


For the results shown in  Fig.~\ref{fig1}, throughout the momentum sphere there are no and thus everywhere $S_3$ is not singular. This is quite expected, as codimension analysis has revealed that  \textbf{V} points are not generic and generally would be absent~\cite{NYE_natural_1999,Supplemental_Material}. That is, extra symmetries are required to achieve strictly zero scattering.  In Fig.~\ref{fig2}(a) we demonstrate scattering by another electromagnetic duality-symmetric (self-dual) cylinder ($\epsilon=\mu=2.25$)~\cite{FERNANDEZ-CORBATON_2013_Phys.Rev.Lett._Electromagnetica,FERNANDEZ-CORBATON_2013_Opt.ExpressOE_Forwarda,YANG_2020_ACSPhotonics_Electromagnetic}. One special property for self-dual particles is: with an incident CP wave, helicity conservation requires that scatterings along all directions are also CP with the same handedness as that of the incident wave~\cite{FERNANDEZ-CORBATON_2013_Phys.Rev.Lett._Electromagnetica,FERNANDEZ-CORBATON_2013_Opt.ExpressOE_Forwarda,YANG_2020_ACSPhotonics_Electromagnetic}.  From the perspective of steady line fields, this scenario is trivial as the field is singular everywhere; while from the perspective of instantaneous vector fields, this is contrastingly nontrivial: as we have revealed, since vector singularities do not locate on  CP positions, Eq.~(\ref{PHtheorem-vector}) secures the existence of at least one dark direction (\textbf{V} point). 

Distributions of  $I_{\mathrm{sca}}$ and instantaneous field vectors for LCP and RCP incidences ($\lambda=810$ nm) are shown respectively in Figs.~\ref{fig2}(b) and \ref{fig2}(c), where in-view \textbf{V} points ($\mathbf{V}_{\mathrm{L}}$ and $\mathbf{V}_{\mathrm{R}}$, respectively) are marked  with integer indexes also specified. Index sum of all \textbf{V} points (including out-of-view ones) is indeed $2$~\cite{Supplemental_Material}. Different from the vector field singularities [shown in  Figs.~\ref{fig1}(d)-\ref{fig1}(f)] that move along lines of linear polarizations, here there are no such lines and all singularities are fixed at \textbf{V} points. This is evident from the insets of  Figs.~\ref{fig2}(d) and \ref{fig2}(e) that show the scattering distributions at another instant: the field vectors do rotate in time while \textbf{V} points are fixed in space. 

Another consequence of helicity conservation is that, along the dark directions of $\mathbf{V}_{\mathrm{L}}$ and $\mathbf{V}_{\mathrm{R}}$, the scattering will remain to be respectively RCP ($S_3=-1$) and LCP ($S_3=1$), irrespective of the incident polarizations. This is easy to comprehend~\cite{Supplemental_Material}: \textbf{(i)} both incident and scattered waves can be expanded into LCP and RCP components; \textbf{(ii)} helicity conservation ensures that there are no cross couplings between them,  incident LCP (RCP) components being scattered into LCP (RCP) components only; \textbf{(iii) } along the scattering direction of $\mathbf{V}_{\mathrm{L}}$ ($\mathbf{V}_{\mathrm{R}}$), for arbitrary incident polarizations, LCP (RCP) components would be absent and thus the scattered waves would invariantly be RCP (LCP). This is manifest Fig.~\ref{fig2}(f) where we demonstrate invariant $S_3$ on a pair of $\mathbf{V}_{\mathrm{L}}$ and $\mathbf{V}_{\mathrm{R}}$  marked in Figs.~\ref{fig2}(d) and \ref{fig2}(e). The incident polarizations evolve on two great circles on the Poincaré sphere [see the inst of Fig.~\ref{fig2}(f)], which are parameterized by $\alpha$ [linear polarizations along \textbf{x} ($\alpha=0$) and \textbf{y} axes ($\alpha=\pi$)] and $\beta$ [LCP ($\beta=0$) and RCP ($\beta=\pi$)], respectively.

\begin{figure}[tp]
\centerline{\includegraphics[width=8cm]{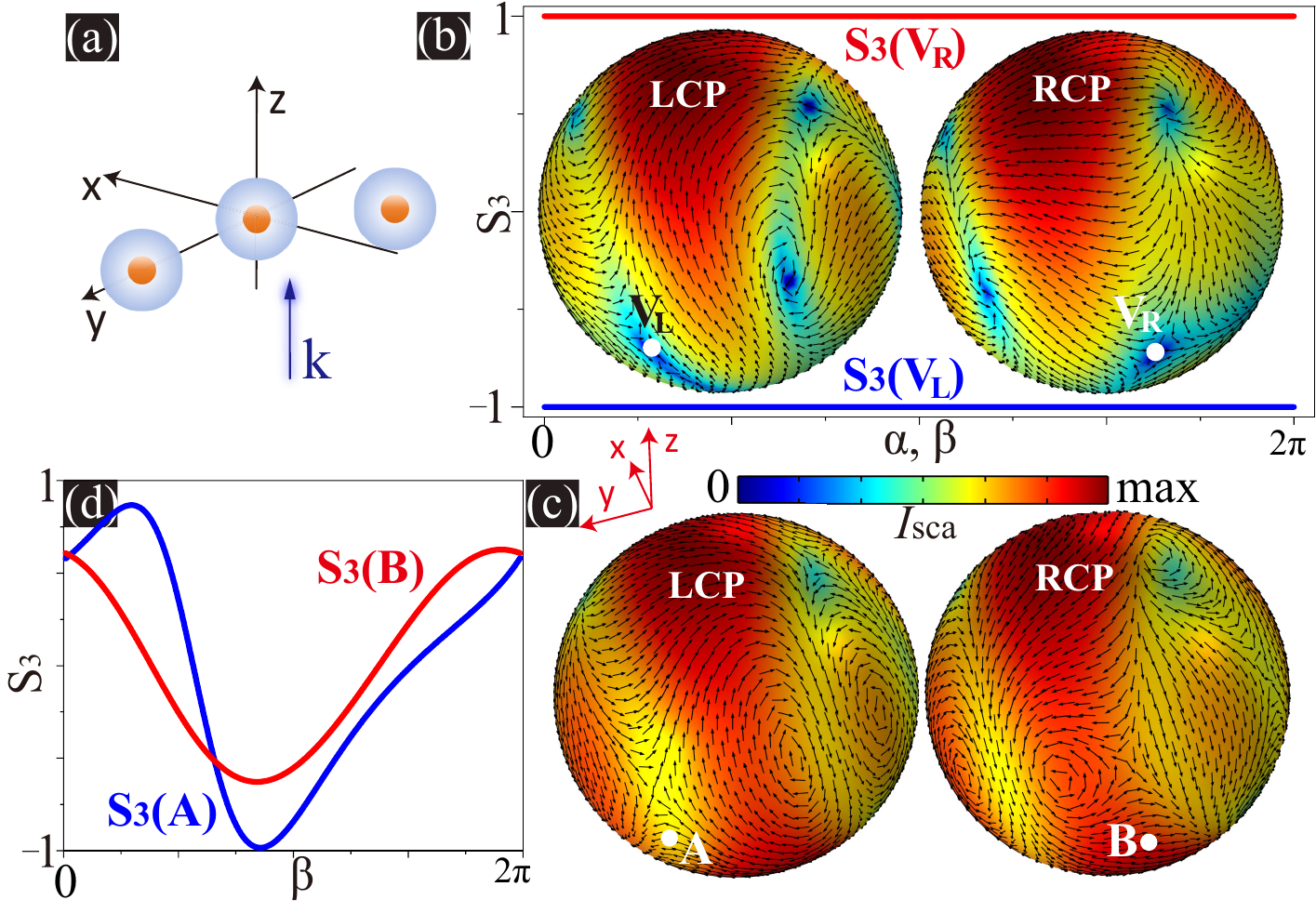}} \caption{\small (a) The scattering of plane waves along +\textbf{z} by three self-dual core-shell particles. The two insets of (b) show  $I_{\mathrm{sca}}$ and field vector distributions respectively for LCP and RCP incidences, with  two \textbf{V} points marked. The invariant scattering polarizations with varying incident polarizations on marked points are shown in (b). Except that the incident wavelength is shifted to make the system not self-dual: (c) is the same as the two insets in (b);  (d) is the same (only $\beta$-dependence is shown) as (b).
\textbf{A} (\textbf{B}) and  $\mathbf{V}_{\mathrm{L}}$ ($\mathbf{V}_{\mathrm{R}}$) correspond to the same direction.}
\label{fig3}
\end{figure}

\begin{figure}[tp]
\centerline{\includegraphics[width=9cm]{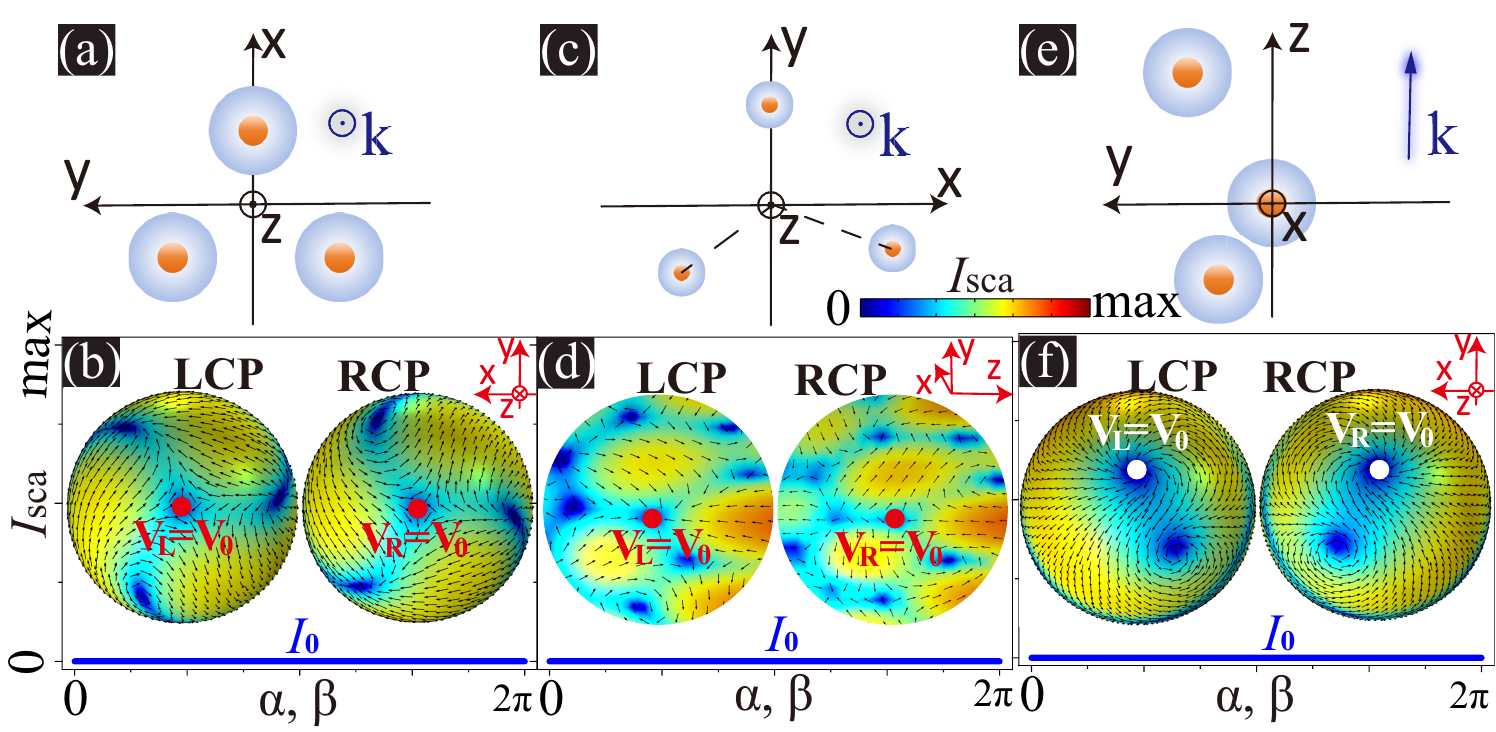}} \caption{\small  Scattering of plane waves along +\textbf{z} by three tri-particle ensembles: (a) the configuration is rotationally symmetric with rotation axis parallel to \textbf{z}-axis;  the inter-particle couplings are weaker in (c) with far-separated particles, and stronger in (e) with closely-packed particles (see \cite{Supplemental_Material} for detailed position information). The corresponding invariant zero scattering with varying incident polarizations along the direction of $\mathbf{V}_0$ (marked in insets that show  $I_{\mathrm{sca}}$ and vector field distributions with CP incidences) is shown in (b), (d) and (f), respectively. For the insets in (d), only field distributions on a spherical disck of the momentum sphere (centred around the direction close to -\textbf{x}) are shown to better visualize the overlapping of $\mathbf{V}_{\mathrm{L}}$ and $\mathbf{V}_{\mathrm{R}}$}
\label{fig4}
\end{figure}

So far the principles we reveal have been demonstrated through a hypothetic self-dual magnetic particle ($\mu=2.25$).  We now turn to practical nonmagnetic self-dual particles, of which the magnetic responses originate from optically-induced magnetism~\cite{Wheeler2006_PRB,Liu2012_ACSNANO,Liu2014_CPB,KUZNETSOV_Science_optically_2016}. For spherical particles that support dominantly lowest-order dipolar responses, the effective electric and magnetic isotropic polarizabilities are~\cite{Wheeler2006_PRB,Liu2012_ACSNANO,Bohren1983_book}: ${\alpha_{e}} = {{3i} \over {2{k^3}}}{a_1},~{\alpha_{m}} = {{3i} \over {2{k^3}}}{b_1}$,
where $a_1$ and $b_1$ are Mie scattering coefficients~\cite{Bohren1983_book}. The spherical particle becomes self-dual when $a_1=b_1$, meaning that electric and magnetic dipolar responses are balanced ${\alpha_{e}} = {\alpha_{m}}$.  Here we employ a similar particle as that adopted in our previous studies ~\cite{Liu2012_ACSNANO,YANG_2020_ACSPhotonics_Electromagnetic}: Ag core-dielectric ($\epsilon=3.4^2$) shell spherical particle with inner radius $R_1=40$~nm and outer radius $R_2=160$~nm; the permittivity of silver is extracted from the experimental data in Ref.~\cite{Johnson1972_PRB}. This particle becomes self-dual at $\lambda_0=1115$~nm only where $a_1 \approx b_1$ (refer to Ref.~\cite{Supplemental_Material} for discussions about non-ideal dual symmetry). Here we study the scattering by a self-dual ensemble consisting of three such particles [Fig.~\ref{fig3}(a); see \cite{Supplemental_Material} for detailed position information, as is also the case for ensembles in Fig.~\ref{fig4}).  Scattering distributions at one instant for $\lambda=\lambda_0$ are shown as insets in Fig.~\ref{fig3}(b), where a pair of \textbf{V} points.  The invariance of $S_3$ on both points with varying incident polarizations is summarized in Fig.~\ref{fig3}(b).  For comparison, we shift the incident wavelength to $\lambda_0=1040$~nm where the system is not self-dual anymore ($a_1 \neq b_1$). As shown in Fig.~\ref{fig3}(c),  along the directions of \textbf{A} and \textbf{B} [identical directions as those of $\mathbf{V}_{\mathrm{L}}$ and $\mathbf{V}_{\mathrm{R}}$ in Fig.~\ref{fig3}(b), respectively],  here the scattering is not zero anymore. 
We further demonstrate the evolution of $S_3$ along those directions in Fig.~\ref{fig3}(d) for different incident polarizations. As expected, $S_3$ varies and are not fixed at $\pm 1$ anymore. 

It is clear from Figs.~\ref{fig2} and ~\ref{fig3} that $\mathbf{V}_{\mathrm{L}}$ and $\mathbf{V}_{\mathrm{R}}$ generally are spatially separated (never located on the forward direction, as required by the optical theorem~\cite{Bohren1983_book}). When they are tuned to overlap at point $\mathbf{V}_{0}=\mathbf{V}_{\mathrm{L}}=\mathbf{V}_{\mathrm{R}}$, on $\mathbf{V}_0$ the scattered field would contain neither LCP nor LCP components, and thus have to be zero for arbitrary incident polarizations.  Such $\mathbf{V}$-point overlapping can be secured by an extra geometric symmetry~\cite{FERNANDEZ-CORBATON_2013_Opt.ExpressOE_Forwarda,YANG_2020_ACSPhotonics_Electromagnetic}: for a self-dual scattering system that exhibits no less than $3$-fold rotation symmetry, with CP waves incident along the rotation axis, $\mathbf{V}_{\mathrm{L}}$ and $\mathbf{V}_{\mathrm{R}}$ would overlap on the backscattering direction. One such scattering configuration is shown in Fig.~\ref{fig4}(a) and $\mathbf{V}_0$ is marked in insets of Fig.~\ref{fig4}(b). The directional scattering intensity on $\mathbf{V}_0$  ($I_0$) is invariantly zero for varying incident polarizations, as verified by Fig.~\ref{fig4}(b).

For an ensemble of randomly distributed such self-dual particles (${\alpha_{e}} = {\alpha_{m}}={\alpha_{0}}$), the scattering can be described by the coupled dipole equation~\cite{Supplemental_Material,Liu2012_ACSNANO,Mulholland1994_Langmuir,Merchiers2007_PRA}. 
When the particles are separated by sufficiently large distances so that the couplings among them are negligible~\cite{Supplemental_Material}, each particle supports a pair of orthogonal and equal electric and magnetic dipolar moments (Kerker particles~\cite{Kerker1983_JOSA,GARCIA-CAMARA_2011_Opt.Lett.OL_Directionality,Geffrin2012_NC,Fu2013_NC,Person2013_NL,ZAMBRANA-PUYALTO2013Opt.Lett.,LIU_2018_Opt.Express_Generalized,OLMOS-TRIGO_2020_Phys.Rev.Lett._Kerker,LASA-ALONSO_2022_Mater.Adv._Correlations}) and the backward scattering is zero for both individual particles and the whole ensemble.  When the couplings among the particles are non-negligible and at the same time the scattering configuration does not exhibit the rotation symmetry required, there is no mechanism to guarantee overlapped $\mathbf{V}_{\mathrm{L}}$ and $\mathbf{V}_{\mathrm{R}}$ at the backward direction or any other directions (see Fig.~\ref{fig3}). Nevertheless, accidental overlapping is still accessible by tuning the incident direction and particle distributions. Two such scattering configurations are shown  in Fig.~\ref{fig4}(c) [Fig.~\ref{fig4}(e)] where the particles are well-separated (closed-packed)~\cite{Supplemental_Material} with weaker (stronger) inter-particle couplings.  For both scenarios, overlapped  $\mathbf{V}_0$  are obtained [marked in the insets of Figs.~\ref{fig4}(d) and \ref{fig4}(f)] and neither locate on the backward direction.  As expected, the directional scattering intensity on $\mathbf{V}_0$ for both scenarios remains to be zero, irrespective of the incident polarizations [Figs.~\ref{fig4}(d) and \ref{fig4}(f)]. In fact, the existence of overlapped $\mathbf{V}_0$ can be verified from a single incidence: when the incident polarization is non-circular (contains both LCP and RCP components), the dark direction must correspond to an overlapped $\mathbf{V}_0$, since the scattered LCP and RCP components do not interfere and thus they both have to be zero~\cite{Supplemental_Material}. That is, for self-dual scattering systems, \textit{{the dark direction for one non-circular incident polarization would remain dark for all incident polarizations}}~\cite{Supplemental_Material}.

In conclusion, it is uncovered that \textit{polarizations underdescribe vectorial electromagnetic waves} and both steady and instantaneous quantities and structures are required to capture the subtleties of electromagnetism. We reveal that the Poincaré-Hopf theorem regulates the distributions of not only steady line fields but also instantaneous vector fields, for which the generic singularities are respectively fixed \textbf{C} points and cyclically moving (along lines of linear polarizations) vector zeros (\textbf{Z} points), with the index sum being identically the Euler characteristic $2$. Besides the global application throughout
the momentum sphere, instantaneous singularities can be also
applied locally to an arbitrarily chosen loop~\cite{Supplemental_Material}. Our discovery is independent of optical or geometric parameters of the scatterers, or structures of incident waves. As a result, our demonstrations with self-dual particles is only a special application of it. The complementarity we reveal might be broadly applied in other branches of wave physics, where singularities are generic and ubiquitous.

\emph{Acknowledgment}:  This research was funded by the National Natural Science
 Foundation of China (Grants No. 12274462, No.11674396, 
 and No. 11874426) and several other projects of Hunan 
 Province (Projects No. 2024JJ2056, No. 2023JJ10051, No. 
 2018JJ1033, and No. 2017RS3039). W. L. acknowledges many illuminating correspondences with Sir Michael Berry, whose monumental paper~\cite{NYE_1974_Proc.R.Soc.Lond.Math.Phys.Sci._Dislocations} with J. F. Nye on phase singularities was published $50$ years ago.



%





\onecolumngrid
\clearpage


{\centering
  \noindent\textbf{\large{Supplemental Material for:}}
\\\vspace{0.1cm}
\noindent\textbf{\large{``Polarizations Underdescribe Vectorial Electromagnetic Waves"}}
\\\bigskip


Chunchao Wen$^{1,2}$,  Jianfa Zhang$^{1,2,*}$, Chaofan Zhang$^{1,2,\dag}$, Shiqiao Qin$^{1,2}$, Zhihong Zhu$^{1,2}$, and Wei Liu$^{1,2,\ddagger}$
\\
\small{$^1$ \emph{College for Advanced Interdisciplinary Studies, National University of Defense Technology, Changsha 410073, P. R. China.}}\\
\small{$^2$ \emph{Nanhu Laser Laboratory and Hunan Provincial Key Laboratory of Novel Nano-Optoelectronic Information Materials and Devices,
National University of Defense Technology, Changsha 410073, P. R. China.}}\\
$^{*}$~jfzhang85@nudt.edu.cn; $^{\dag}$~c.zhang@nudt.edu.cn; $^{\ddagger}$~wei.liu.pku@gmail.com\\ \vspace{0.3cm}

{\leftskip=0pt \rightskip=0pt plus 0cm

The Supplemental Material includes the following eleven sections: (\textbf{\uppercase\expandafter{\romannumeral1}}). Cartesian coordinates of consisting core-shell particles for the ensembles studied;
(\textbf{\uppercase\expandafter{\romannumeral2}}). All line-field singularities in Figs. 2(b) and 2(c) and their indexes;
(\textbf{\uppercase\expandafter{\romannumeral3}}). All instantaneous vector-field singularities in  Fig. 2(f) and their indexes; 
(\textbf{\uppercase\expandafter{\romannumeral4}}). Cyclic movement of instantaneous vector-field singularities along lines of linear polarizations in Figs.2(d)-2(f);
(\textbf{\uppercase\expandafter{\romannumeral5}}). All \textbf{V} points in Figs. 3(b) and 3(c) and their indexes;
(\textbf{\uppercase\expandafter{\romannumeral6}}). Formalisms for scatterings by self-dual particles;
(\textbf{\uppercase\expandafter{\romannumeral7}}). Effect of non-ideal self-duality;
(\textbf{\uppercase\expandafter{\romannumeral8}}). Coupled dipole theory for self-dual dipolar particles;
(\textbf{\uppercase\expandafter{\romannumeral9}}). Local singularities and its connection with global topology through the Poincar$\mathrm{\acute{{e}}}$-Hopf theorem;
(\textbf{\uppercase\expandafter{\romannumeral10}}). Local applications of instantaneous singularities;
(\textbf{\uppercase\expandafter{\romannumeral11}}). Codimension analysis for \textbf{V} points.\\\bigskip 

\setcounter{equation}{0}
\setcounter{figure}{0}
\newcounter{sfigure}
\setcounter{sfigure}{1}
\setcounter{table}{0}
\renewcommand{\theequation}{S\arabic{equation}}

 \renewcommand\thefigure{S{\arabic{figure}}}
\renewcommand{\thesection}{S\arabic{section}}

\twocolumngrid
\section{(\textbf{\uppercase\expandafter{\romannumeral1}}).  Cartesian coordinates of consisting core-shell particles for the ensembles studied.}

In Table (\ref{table}) we have summarized the Cartesian coordinates of all consisting core-shell particles centered at $\mathbf{r}_i$ for the ensembles studied in both Fig. 4 and Fig. 5.

\begin{table}[htp]
\centering 
\begin{tabular}{|l|c|c|c|}
\hline & $\mathbf{r}_1/\mu$m&  $\mathbf{r}_2/\mu$m  &  $\mathbf{r}_3/\mu$m  \\
\hline Fig. 4 (a) & $(0,0.72,0)$ & $(0,0,0)$ & $(-0.5,-0.3,0)$ \\
\hline Fig. 5 (a) & $(-0.144,0.25,0)$ & $(-0.144,-0.25,0)$ & $(0.289,0,0)$ \\
\hline Fig. 5 (c) & $(-1.8,-1.8,0)$ & $(2.511,-0.443,0)$ & $(0,2.55,0)$ \\
\hline Fig. 5 (e) & $(0,0.283,-0.510)$ & $(0,0,0)$ & $(0,0.36,0.624)$ \\
\hline
\end{tabular}
\caption{Cartesian coordinates of consisting core-shell particles for the ensembles studied in Fig. 4 and Fig. 5 in the Main Letter.} 
\label{table} 
\end{table}

\section{(\textbf{\uppercase\expandafter{\romannumeral2}}).  All line-field singularities in Figs. 2(b) and 2(c) and their indexes}
In Fig. \ref{figure1_SP} we have shown and marked all line-field singularities (\textbf{C} points) and specified their indexes for the scattering configuration shown in Fig. 2(a), with LCP incidence. Four singularities are shown already in Figs. 2(b) and 2(c). The index sum of all six singularities are  $5\times(+\frac{1}{2})+ 1\times(-\frac{1}{2})=\chi=2$.

\section{(\textbf{\uppercase\expandafter{\romannumeral3}}). All instantaneous vector-field singularities in  Fig. 2(f) and their indexes}

In Fig. \ref{figure2_SP} we have shown and marked all instantaneous vector-field singularities [\textbf{Z} points; at the same instant as that in Fig. 2(f)] and specified their indexes for the scattering configuration shown in Fig. 2(a), with LCP incidence.  Two singularities are shown already in Fig. 2(f). The index sum of all four singularities are  $3\times(+1)+ 1\times(-1)=\chi=2$.

\begin{figure}[htbp]
\centerline{\includegraphics[width=8.5cm]{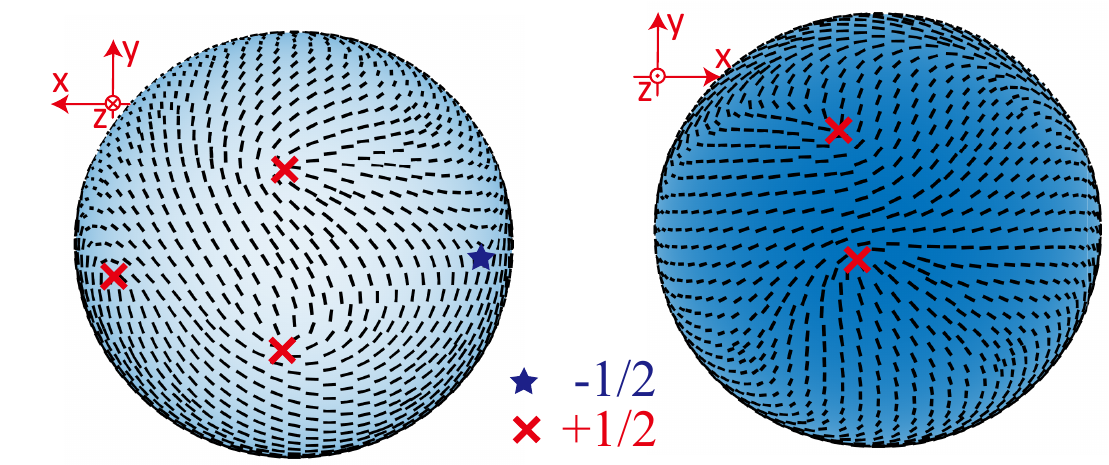}} \caption{\small Polarization ($S_3$) distributions of scattered fields on the momentum sphere as viewed from different angles. All  line-field singularities are marked with their indexes specified. The scattering configuration is shown in Fig. 2 (LCP incidence), and four singularities are already shown in  Figs. 2(b) and 2(c).}
\label{figure1_SP}
\end{figure}

\begin{figure}[htbp]
\centerline{\includegraphics[width=8.5cm]{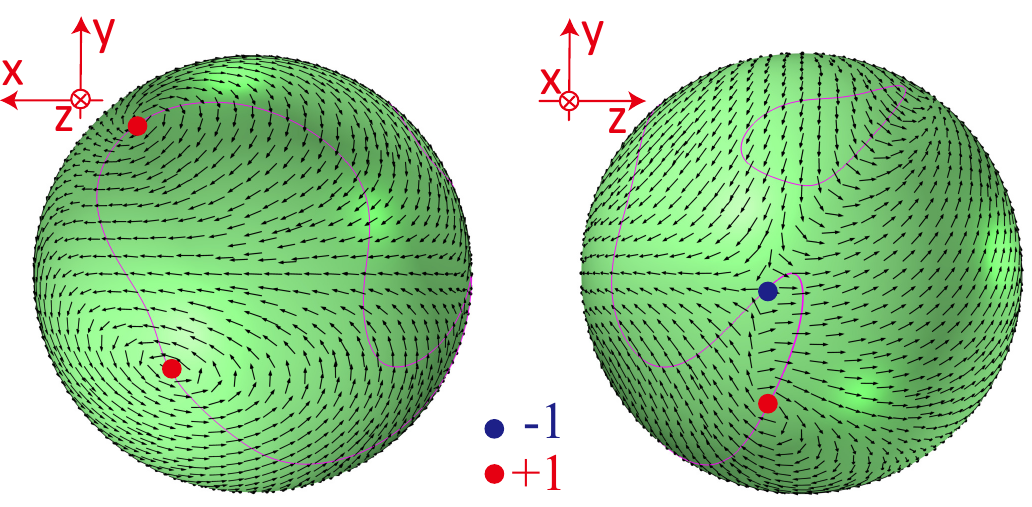}} \caption{\small Instantaneous vector-field distributions of scattered fields on the momentum sphere as viewed from different angles [at the same instant as that in Fig. 2(f)]. All vector-field singularities are marked with their indexes specified. The scattering configuration is shown in Fig. 2 (LCP incidence), and two singularities are already shown in  Fig. 2(f).}
\label{figure2_SP}
\end{figure}

\section{(\textbf{\uppercase\expandafter{\romannumeral4}}).  Cyclic movement of instantaneous vector-field singularities along lines of linear polarizations in Figs.2(d)-2(f).}
In Fig. \ref{figure3_SP} we show more detailed evolutions of the instantaneous vector-field singularities (\textbf{V} points) along lines of linear polarizations in a full cycle $\mathrm{T}=\lambda/c$. The distributions at three different instants are already shown in Figs.2(d)-2(f). more details about the birth and annihilation of opposite-index singularities on lines of linear polarizations could be found in the supplemental video.

\begin{figure}[htbp]
\centerline{\includegraphics[width=8.5cm]{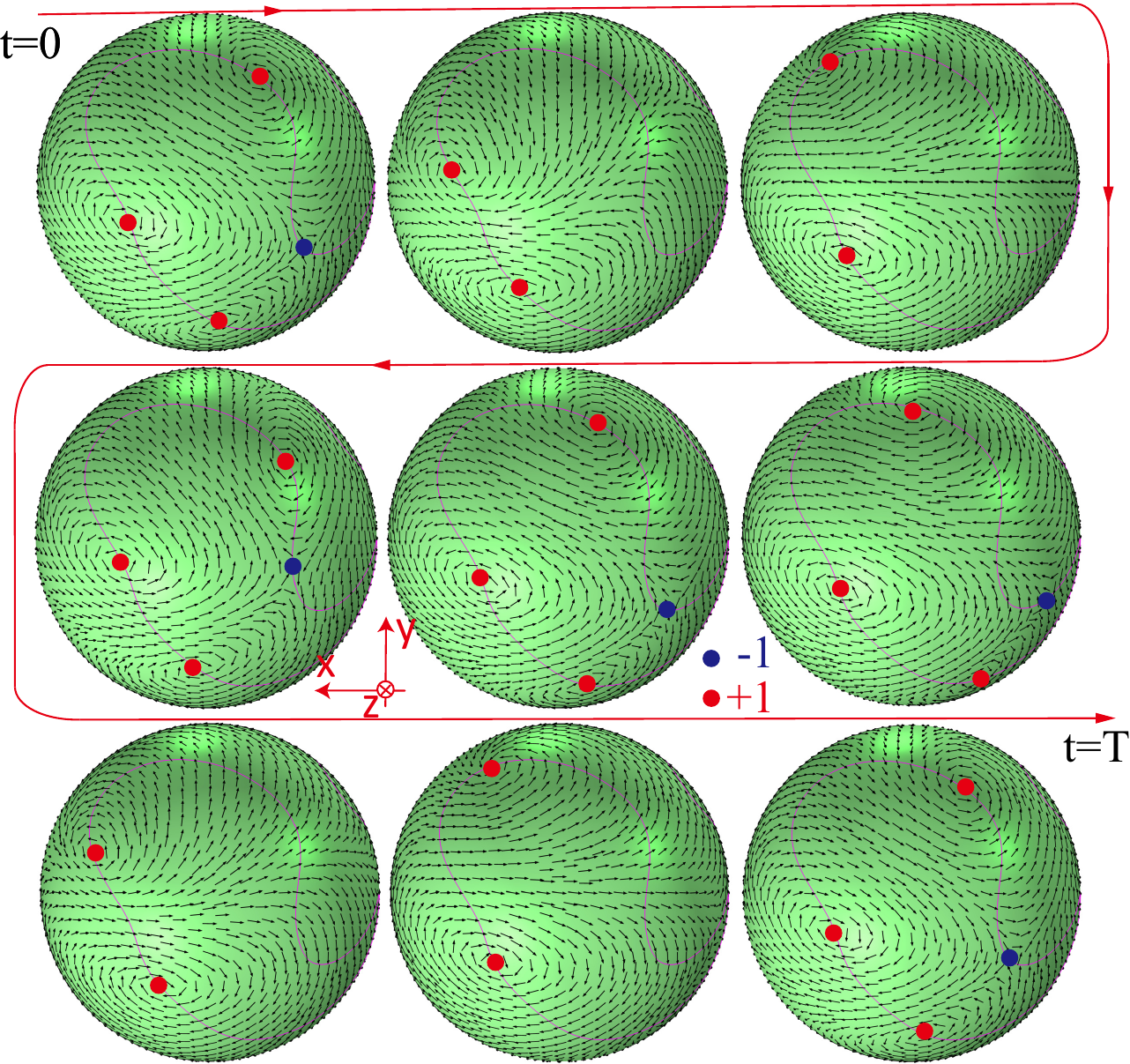}} \caption{\small Cyclic movement of instantaneous vector-field singularities along lines of linear polarizations in a full cycle $\mathrm{T}$. The scattering configuration is shown in Fig. 2(a) with LCP incidence, and field distribuitons at three instants are already shown in  Figs.2(d)-2(f).}
\label{figure3_SP}
\end{figure}

\begin{figure}[htbp]
\centerline{\includegraphics[width=6.5cm]{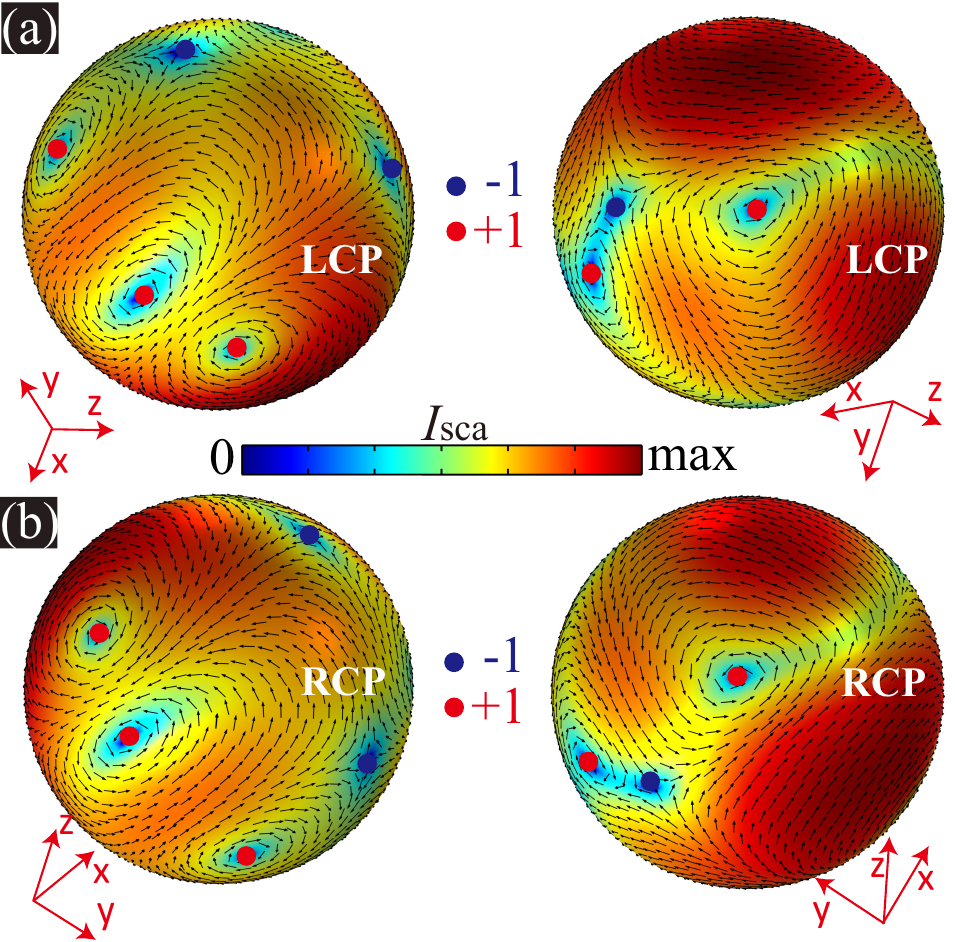}} \caption{\small Distributions of \textbf{V} points for scattered fields on the momentum sphere as viewed from different angles, at the same instant (a) as that in Fig. 3(b) for LCP incidence, and (b) as that in Fig. 3(c) for RCP incidence. All vector-field singularities are marked with their indexes specified. The scattering configuration is shown in Fig. 3(a), and five singularities are already shown in both Figs. 3(b) and 3(c).}
\label{figure4_SP}
\end{figure}

\section{(\textbf{\uppercase\expandafter{\romannumeral5}}).  All dark directions (\textbf{V} points) in Figs. 3(b) and 3(c) and their indexes.}

In Fig. \ref{figure4_SP} we have shown and marked all dark directions (\textbf{V}) points) and specified their indexes for the scattering configuration shown in Fig. 3(a) with both LCP and RCP incidences. Five of the \textbf{V}) points are shown already in both Figs.3(b) and 3(c), for LCP and RCP incidences respectively.  For both scenarios in Figs. \ref{figure4_SP}(a) and  \ref{figure4_SP}(b), the index sum of all $8$ singularities are  $5\times(+1)+ 3\times(-1)=\chi=2$.\\

\section{(\textbf{\uppercase\expandafter{\romannumeral6}}).  Formalisms for scatterings by self-dual particles.}

An arbitrarily polarized incident plane wave (denoted by $\mathbf{E}_i$)  can be expressed (in circular basis $\hat{\mathbf{{{L}}}}$ and $\hat{\mathbf{{{R}}}}$) as:
\begin{equation}
\mathbf{E}_i=\gamma_\mathbf{\mathrm{L}} \mathbf{{\hat{L}}}+\gamma_\mathbf{\mathrm{R}}\mathbf{{\hat{R}}},
\end{equation}
where the amplitude and phase of ${\gamma_\mathbf{\mathrm{L}}}/{\gamma_\mathbf{\mathrm{R}}}$ decide  the ellipticity and orientation direction of the incident polarization ellipse, respectively~\cite{YARIV_2006__Photonics}. When the scatterers are self-dual, the helicity conservation ensures that  incident LCP (RCP) components being scattered into LCP (RCP) components of the scattered field only~\cite{FERNANDEZ-CORBATON_2013_Phys.Rev.Lett._Electromagnetica,YANG_2020_ACSPhotonics_Electromagnetic}.  Then scattered waves in the far field (denoted by $\mathbf{E}_s$) along the unit direction vector $\mathbf{\hat{r}}$ can be expressed as:
\begin{equation}
\label{scattering-sm}
\mathbf{E}_s(\mathbf{\hat{r}})=\hat{\mathbf{S}}(\mathbf{\hat{r}}) \mathbf{E}_{i}=\gamma_\mathbf{\mathrm{L}} s_\mathbf{\mathrm{L}}\mathbf{{\hat{L}}}+\gamma_\mathbf{\mathrm{R}}s_\mathbf{\mathrm{R}}\mathbf{{\hat{R}}},
\end{equation}
where $\hat{\mathbf{S}}$ is the scattering matrix~\cite{Bohren1983_book}; $\hat{\mathbf{S}}(\mathbf{\hat{r}}) \mathbf{{\hat{L}}}=s_\mathbf{\mathrm{L}}\mathbf{{\hat{L}}}$; $\hat{\mathbf{S}}(\mathbf{\hat{r}}) \mathbf{{\hat{R}}}=s_\mathbf{\mathrm{R}}\mathbf{{\hat{R}}}$. That is, complex $s_\mathbf{\mathrm{L}}$ ($s_\mathbf{\mathrm{R}}$) characterizes the directional (along $\mathbf{\hat{r}}$) scattering amplitude and phase for LCP (RCP) incidence, respectively~\cite{Bohren1983_book}.  As a result, on $\mathbf{V}_{\mathrm{L}}$ and $\mathbf{V}_{\mathrm{R}}$, we have respectively $s_\mathbf{\mathrm{L}}=0$ and $s_\mathbf{\mathrm{R}}=0$; On the overlapped $\mathbf{V}_{0}$, we have $s_\mathbf{\mathrm{L}}=s_\mathbf{\mathrm{R}}=0$.

According to Eq.~(\ref{scattering-sm}), for arbitrary incident polarizations: (i) on $\mathbf{V}_{\mathrm{L}}$, $s_\mathbf{\mathrm{L}}=0$  (on $\mathbf{V}_{\mathrm{R}}$, $s_\mathbf{\mathrm{R}}=0$), and the scattered field would be RCP (LCP), with $S_3=-1$ ($S_3=1$), which directly explains the results shown in Fig. 4(e) and Fig. 5(b); (ii)   on  $\mathbf{V}_{0}$,  $s_\mathbf{\mathrm{L}}=s_\mathbf{\mathrm{R}}=0$, and thus the scattered field is invariantly zero $\mathbf{E}_s(\mathbf{\hat{r}})=0$ for arbitrary incident polarizations, as is shown in Fig. 5.

For the scattered field expressed by Eq.~(\ref{scattering-sm}), the corresponding scattering intensity is:
\begin{equation}
\label{intensity-sm}
I_s(\mathbf{\hat{r}})=|\mathbf{E}_s(\mathbf{\hat{r}})|^2=|\gamma_\mathbf{\mathrm{L}}|^2 |s_\mathbf{\mathrm{L}}|^2+|\gamma_\mathbf{\mathrm{R}}|^2|s_\mathbf{\mathrm{R}}|^2.
\end{equation}
According to Eq.~(\ref{intensity-sm}), for a non-circular incident polarization (${\gamma_\mathbf{\mathrm{L}}}\neq0$ and ${\gamma_\mathbf{\mathrm{R}}}\neq0$), the existence of a dark direction requires that $s_\mathbf{\mathrm{L}}=s_\mathbf{\mathrm{R}}=0$ ($\mathbf{V}_{\mathrm{L}}$ and $\mathbf{V}_{\mathrm{R}}$ overlapped).  This proves the principle we reveal in the Main Letter: \textit{{Concerning self-dual scattering systems, the dark direction for one non-circular incident polarization would remain dark for all incident polarizations. }}

\section{(\textbf{\uppercase\expandafter{\romannumeral7}}).  Effect of non-ideal self-duality.}

Whether or not the scattering intensity is perfectly (rigorously) zero is an interesting mathematical question but it is of limited significance in physics. In a real physical world, we can never decide rigourously the geometric or physical parameters of the scatterers, and neither can our detectors (or numerical simulators based on  finite-element simulations) reach infinitely small resolutions to decide if the zero is rigourous. Moreover, in the deeper quantum world, rigorous zeros in classical physics would be removed by quantum fluctuations and other quantum effects~\cite{BERRY_2004_J.Opt.A:PureAppl.Opt._Quantum}.

For classical electromagnetic scattering, codimension analysis~\cite{BERRY_2004_J.Opt.PureAppl.Opt._Index} reveals that dark directions on the momentum sphere are not generic: they would be broken into pairs of C points (directions of circular polarizations) upon arbitrarily small perturbations (\textit{e.g.} fabrication imperfections)~\cite{NYE_natural_1999,CHEN_2019_ArXiv190409910Math-PhPhysicsphysics_Linea}.  Nevertheless, the scattering intensity and angular separation between C points (which is zero at dark directions) only changes continuously and thus for tiny perturbations it would remain close to be zero that is beyond the (angular) resolutions of the physical detectors and numerical simulators. From this perspective, previous claims of experimentally (numerically) observed zero scattering could be mostly incorrect mathematically (due to inevitable tiny perturbations) while could be valid physically (due to finite resolutions of detectors and simulators). 

The core-shell spherical particles we employ in Figs. 4 and 5 are not ideally self-dual ($a_1\approx b_1$; $a_n,b_n\approx0$ for $n>2$).  As a result, the dark directions pinpointed in Figs. 4 and 5 are probably not rigourously dark; we mark them as dark since scattering intensities at those positions go asymptotically to zero as we reduce the simulation mesh sizes.  

The validity of our conclusion has nothing to do with the optical or geometric parameters of the scatterers (including electromagnetic duality symmetry), and thus our demonstration with self-dual particles is only a special application of our discovery.  For other non-self-dual scatterers, if we know that throughout the momentum sphere scatterings of linear polarizations  are forbidden ($S_3\neq0$;  self-dual scatterers with circular polarization incidences is a special scenario of $S_3=\pm1$), our discovery immediately tells that there must exists dark directions. 

\section{(\textbf{\uppercase\expandafter{\romannumeral8}}).  Coupled dipole theory for  self-dual dipolar particles.}

For an ensemble of randomly distributed self-dual dipolar particles (${\alpha_{e}} = {\alpha_{m}}={\alpha_{0}}$), the scattering can be described by the coupled dipole theory:
\begin{equation}
\label{coupled-dipole}
\begin{aligned}
&\mathbf{p}_{i}=\alpha_{0} \mathbf{E}_{i}^{0}+\alpha_{0} \sum_{j \neq i}\left(\mathbf{E}_{i}^{\mathbf{p}_{j}}+\mathbf{E}_{i}^{\mathbf{m}_{j}}\right);\\
&\mathbf{m}_{i}=\alpha_{0} \mathbf{H}_{i}^{0}+\alpha_{0} \sum_{j \neq i}\left(\mathbf{H}_{i}^{\mathbf{p}_{j}}+\mathbf{H}_{i}^{\mathbf{m}_{j}}\right),
\end{aligned}
\end{equation}
where $\mathbf{p}_i$ and $\mathbf{m}_i$ are electric and magnetic dipolar moments supported by the $i^{th}$ particle centred at $\mathbf{r}_i$, respectively; $\mathbf{E}_i^0$ and $\mathbf{H}_i^0$ are the incident electric and magnetic fields at $\mathbf{r}_i$, respectively; $\mathbf{E}_i^{{\mathbf{p}_j}}$  ($\mathbf{E}_i^{{\mathbf{p}_j}}$) and $\mathbf{H}_i^{{\mathbf{p}_j}}$ ($\mathbf{H}_i^{{\mathbf{m}_j}}$) are electric and magnetic fields radiated by the electric dipole $\mathbf{p}_j$ (magnetic dipole $\mathbf{m}_j$) at $\mathbf{r}_i$, respectively. When the particles are separated by sufficiently large distances so that the couplings among them are negligible $|\mathbf{E}_i^{{\mathbf{p}_j}}|\ll|\mathbf{E}_i^0|$ and $|\mathbf{H}_i^{{\mathbf{p}_j}}|\ll|\mathbf{H}_i^0|$ for all $i$ and $j$, Eq.~({\ref{coupled-dipole}) is reduced to $|\mathbf{p}_{i}|=|\mathbf{m}_{i}|$ and $\mathbf{p}_{i}\perp\mathbf{m}_{i}$ for all $i$. That is, for incident plane wave of arbitrary polarization, each particle supports a pair of orthogonal and equal electric and magnetic dipolar moments (Kerker particles) and the backward scattering is zero for both individual particles and the whole ensemble. 

\section{(\textbf{\uppercase\expandafter{\romannumeral9}}).  Local singularities and its connection with global topology through the Poincar$\mathrm{\acute{\mathbf{e}}}$-Hopf theorem.}

As has been shown in Fig. (1) in the main letter, electromagnetic waves can be described either through instantaneous vector fields or steady line fields (extracted from steady polarization ellipses). Instantaneous vector-field singularities are denoted as \textbf{Z} points and line-field singularities as \textbf{C} points. The positions where the fields are zero at any instant (\textbf{V} points) are singularities of singularities: they can be viewed as both \textbf{Z} points and \textbf{C} points. Several lowest-order vector-field and line-field singularities~\cite{NYE_natural_1999} are shown respectively in Figs. \ref{figure-reply-c0}(a) and \ref{figure-reply-c0}(b). For the source singularity and lemon singularity, a loop is indicated, with the definition of their index explicitly shown  in Figs. \ref{figure-reply-c0}(c) and \ref{figure-reply-c0}(d), respectively. As we go around the loop counterclockwisely from point A back to A through BCDE,  if the fields on those points rotate counterclockwisely (clockwisely), the index of the enclosed singularity is positive (negative), with the absolute value of the index being the rotation angle of the field divided by $2\pi$. For the source and lemon singularities, the field counterclockwise rotation angles are $2\pi$ and $\pi$, making indexes of $+1$ and $+1/2$, respectively. As is clearly shown, there is a profound difference between vector and line fields: for the former the field rotation angle has to been an integer number of $2\pi$, since the fields lines with arrows and only an integer number of $2\pi$ rotation can bring to field back to itself as we go back to the starting point (from point A back to A); while for the latter,  the fields are lines without arrows, an integer number of $\pi$ rotations is sufficient to bring the field back to itself.  This is exactly the reason why the indexes of vector-field and line-field singularities are integers and half-integers, respectively.  

For both vector and line fields defined on a parameter space, there is a connection between index sum of local singularities and global topology of the parameter space, as revealed by the Poincar$\mathrm{\acute{{e}}}$-Hopf theorem~\cite{NEEDHAM__Visuala}. In our study the parameter space is the momentum sphere with Euler characteristic $\chi=2$, for both vector fields and line fields  defined on it, the Poincar$\mathrm{\acute{{e}}}$-Hopf theorem requires that: 
\begin{equation}
\begin{aligned}
& \sum_i \operatorname{Ind}\left(\mathbf{C}_i\right)=2, \\
& \sum_i \operatorname{Ind}\left(\mathbf{Z}_i\right)=2.
\end{aligned}
\end{equation}
To exemplify this connection, we show in  Fig. \ref{figure-reply-c1} the two elementary field distributions on the sphere: for the vector fields in Fig. \ref{figure-reply-c1}(a), there are two center singularities of index $+1$, making the total index of $+2$; for the line fields in in Fig. \ref{figure-reply-c1}(b), there are four lemon singularities (only two are in view) of index $+1/2$, making also the total index of $+2$. Actually, for any continuous fields on the momentum sphere, independent of the number and local distributions of the singularities, their index sum has to be the Euler characteristic $\chi=2$. This is exactly the mathematical foundation based on which our framework incorporating instantaneous singularities is built.

\begin{figure}[htbp]
\centerline{\includegraphics[width=8.9cm]{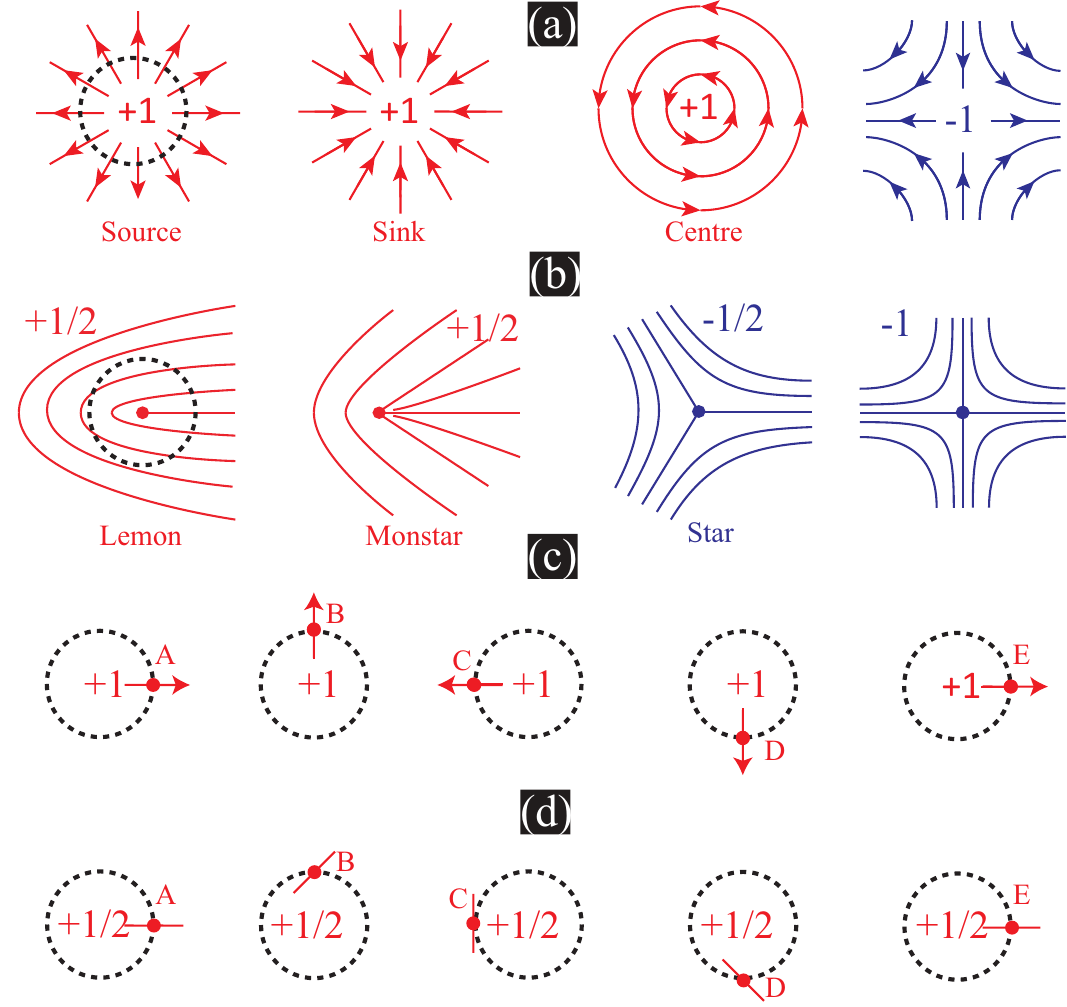}} \caption{\small Several lowest-order vector-field singularities in (a) and line-field singularities in (b). For the source singularity and lemon singularity a loop is indicated, with the definition of their index shown explicitly  in (c) and (d), respectively. Transversing the loop from point A to B, and back to A counterclockwisely,  if the field on those points rotate counterclockwisely (clockwisely), the index of the corresponding singularity is positive (negative), with the absolute value being the rotation angle of the field divided by $2\pi$. For the source and lemon singularities, the counterclockwise field rotation angle is $2\pi$ and $\pi$, making indexes of $+1$ and $+1/2$, respectively.}
\label{figure-reply-c0}
\end{figure}

\begin{figure}[htbp]
\centerline{\includegraphics[width=6cm]{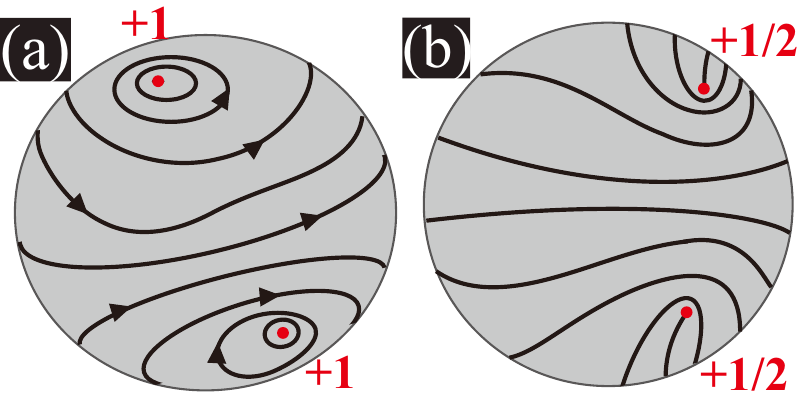}} \caption{\small (a) Vector fields on a sphere with two center singularities of index $+1$. (b) Line fields on a sphere with four lemon singularities (only two are in view) of index $+1/2$. For both cases the index sum of singularities is $+2$, being equal to the Euler characteristic.}
\label{figure-reply-c1}
\end{figure}

\section{(\textbf{\uppercase\expandafter{\romannumeral10}}).  Local applications of instantaneous singularities.}

In the main letter, we apply our framework globally to the whole momentum sphere, and reveal that throughout the momentum sphere if there are no directions of linear polarizations, there must be dark directions of zero scattering. Our framework can also be applied locally to guide the precise identification of the dark directions.  As is shown in Fig. \ref{figure-reply-c2} below, we can select a  closed loop $\mathbb{L}$ (not only applicable for the scattering problem on the momentum sphere,  but also applicable for general electromagnetic problems in other parameter spaces such as the real space). For any instant, if there is no singularity located on $\mathbb{L}$, we can calculate its index (see the definition in the last section and Figs. \ref{figure-reply-c0}) through rotations of instantaneous vectors defined on it [for both Figs. \ref{figure-reply-c2}(a) and \ref{figure-reply-c2}(b) the index for the loop is $\mathbf{Ind}(\mathbb{L})=1$].  For an chosen $\mathbb{L}$  at any instant (for continuous vector fields, the index of the loop  has to be equal to the index sum of all singularities enclosed by it: $\sum_i \operatorname{{\mathbf{Ind}}}({\mathrm{\mathbf{Z}}}_i)=\mathbf{Ind}(\mathbb{L})$~\cite{NEEDHAM__Visuala}): (a) If $\mathbf{Ind}(\mathbb{L})\neq0$ and inside $\mathbb{L}$  there are no linear polarizations, we know that inside $\mathbb{L}$ there must be dark (\textbf{V}) points to account for the nonzero loop index [see Fig. \ref{figure-reply-c2}(a)]; (b) If there are linear polarizations inside $\mathbb{L}$ but index sum of singularities on them  $\sum_i \operatorname{{\mathbf{Ind}}}({\mathrm{\mathbf{Z}}}_i^\mathrm{L})\neq\mathbf{Ind}(\mathbb{L})$, we also know that inside $\mathbb{L}$ there must be dark points to account for the index difference $\sum_i \operatorname{{\mathbf{Ind}}}({\mathrm{\mathbf{V}}}_i)=\mathbf{Ind}(\mathbb{L})-\sum_i \operatorname{{\mathbf{Ind}}}({\mathrm{\mathbf{Z}}}_i^\mathrm{L})$ [see Fig. \ref{figure-reply-c2}(b)]. Here ${\mathrm{\mathbf{Z}}}_i^{\mathrm{L}}$ denote instantaneous singularities on the linear polarizations, and thus do not include dark  points. In Fig. \ref{figure-reply-c2}(c) we showcase a specific vector field with several loops chosen: $\mathbf{Ind}(\mathbb{L})\neq0$ and $\mathbf{Ind}(\mathbb{L})=0$ for blue and red dashed loops, respectively.  It is worth mentioning that for a loop $\mathbf{Ind}(\mathbb{L})=0$,  there could be no singularities inside (smaller red dashed loop) or several singularities with indexes cancelling each other (larger red dashed loop). 

\begin{figure}[htbp]
\centerline{\includegraphics[width=7cm]{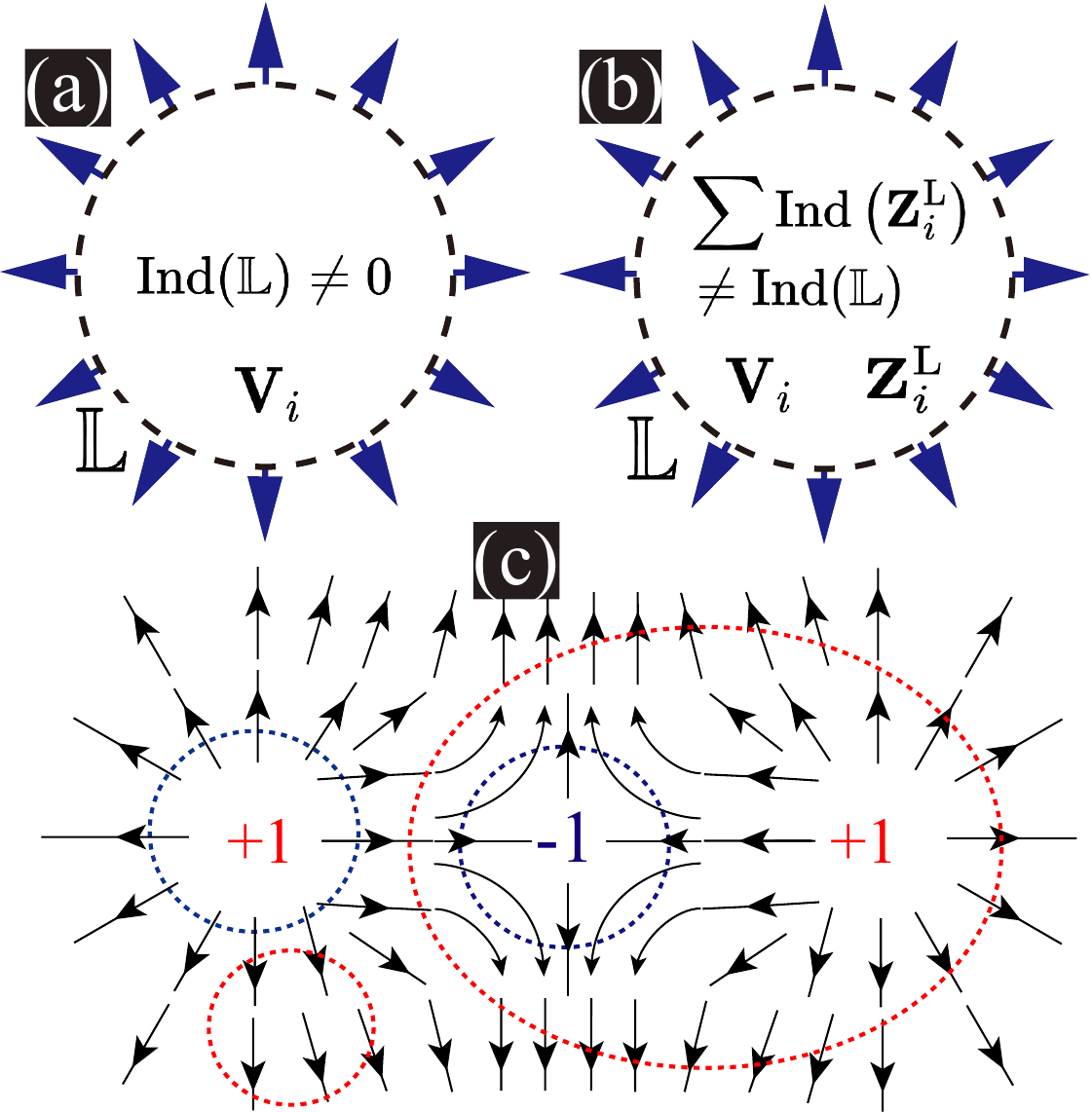}} \caption{\small The existence of \textbf{V} points within the loop $\mathbb{L}$ is secured in both (a) and (b). In (a): $\mathbf{Ind}(\mathbb{L})\neq0$ and inside $\mathbb{L}$ there are no linear polarizations (thus also no ${\mathrm{\mathbf{Z}}}_i^\mathrm{L}$). In (b): though ${\mathrm{\mathbf{Z}}}_i^\mathrm{L}$ is present, $\sum_i \operatorname{{\mathbf{Ind}}}({\mathrm{\mathbf{Z}}}_i^\mathrm{L})\neq\mathbf{Ind}(\mathbb{L})$. (c) A vector field with several loops chosen: $\mathbf{Ind}(\mathbb{L})\neq0$ and $\mathbf{Ind}(\mathbb{L})=0$ for blue and red dashed loops, respectively. }
\label{figure-reply-c2}
\end{figure}

Since the size of $\mathbb{L}$ can be chosen arbitrarily small, a combination of global and local applications of our framework will serve as a powerful guide for precise local identifications of dark points.  For example, the principle underlying Fig. \ref{figure-reply-c2}(a) is directly applicable to the self-dual particle scattering problem: for an arbitrary $\mathbb{L}$ chosen on the momentum sphere with index $\mathbf{Ind}(\mathbb{L})$, since there are no linear polarizations (and thus no ${\mathrm{\mathbf{Z}}}_i^\mathrm{L}$) within any loop $\mathbb{L}$ (non-circular polarizations are forbidden by duality), if $\mathbf{Ind}(\mathbb{L})\neq0$ we can deduce that there must be dark points enclosed. This would be very helpful for more precise local identifications of dark directions.

\section{(\textbf{\uppercase\expandafter{\romannumeral11}}).  Codimension analysis for \textbf{V} points.}

In the far field on the momentum sphere where the electromagnetic field is transverse, the codimension of \textbf{V} point is four~\cite{NYE_natural_1999}, which means that the existence of \textbf{V} point requires all four field components (real and imaginary parts of two transverse orthogonal fields, \textit{e.g.} $\mathbf{E}_x$ and $\mathbf{E}_y$). Nevertheless, the parameter space (momentum space) is only two-dimensional, which means there are only two parameters that can be freely tuned, while to obtain $\mathbf{V}$ points four free parameters are required. As a result, $\mathbf{V}$ points on the momentum sphere are non-generic (accidental) and would be broken into pairs of $\mathbf{C}$ points under perturbations~\cite{NYE_natural_1999}. To protect the existence of \textbf{V} points, extra symmetries are required~\cite{NYE_natural_1999}, and in our demonstrations it is electromagnetic duality symmetry employed.

\end{document}